\documentclass{IEEEtran}
\usepackage{graphicx}
\usepackage{caption}

\usepackage{fixltx2e}
\usepackage{url}

\newcommand{\shorttitle}[1]%
{\markboth{Proceedings of the 1\MakeLowercase{$^{st}$} ICST, Greece 2009}{#1} }
\newcommand{\etal}{\MakeLowercase{\textit{et al. }}} % "et al."

\begin{document}
\title{Indian Payloads (RT-2 Experiment) Onboard CORONAS-PHOTON Mission}

%\author{\IEEEauthorblockN{Anuj Nandi}
%\IEEEauthorblockA{Indian Centre for Space Physics \\
%43-Chalantika, Garia Station Road\\
%Garia, Kolkata, India, 700084\\
%Email: anuj@csp.res.in \\
%Also at: ISRO-HQ, Bangalore, India}
%\and
%\IEEEauthorblockN{A. R. Rao}
%\IEEEauthorblockA{Tata Institute of Fundamental Research\\
%Homi Bhaba Road, Colaba, Mumbai\\
%India, 400005\\
%Email: arrao@tifr.res.in \\
%(PI of the RT-2 Project)}
%\and
%\IEEEauthorblockN{Sandip K. Chakrabarti}
%\IEEEauthorblockA{S. N. Bose National Centre for \\Basic Sciences\\
%Block-JD, Sector-III, Salt Lake\\
%Kolkata, India, 700098\\
%Email: chakraba@bose.res.in \\
%Also at: Indian Centre for Space Physics, \\Kolkata, India\\
%(Co-PI of the Project)}}

\author{\IEEEauthorblockN{Anuj Nandi\IEEEauthorrefmark{1}\IEEEauthorrefmark{5}, 
A. R. Rao\IEEEauthorrefmark{2},
Sandip K. Chakrabarti\IEEEauthorrefmark{3}\IEEEauthorrefmark{1}, J. P. Malkar\IEEEauthorrefmark{2}, 
S. Sreekumar\IEEEauthorrefmark{4}, Dipak Debnath\IEEEauthorrefmark{1}, \\
M. K. Hingar\IEEEauthorrefmark{2}, Tilak Kotoch\IEEEauthorrefmark{1}, Yuri Kotov\IEEEauthorrefmark{6},
Andrey Arkhangelskiy\IEEEauthorrefmark{6}}\\
\IEEEauthorblockA{\IEEEauthorrefmark{1} Indian Centre for Space Physics, 43-Chalantika, Garia Station
Road, Kolkata - 700084, India.\\ Email: anuj@csp.res.in}

\IEEEauthorblockA{\IEEEauthorrefmark{2}Tata Institute of Fundamental Research, Homi Bhaba Road, Colaba,
Mumbai - 400005, India. \\ Email: arrao@tifr.res.in}

\IEEEauthorblockA{\IEEEauthorrefmark{3}S. N. Bose National Centre for Basic Sciences, Block-JD, 
Sector-III, Salt Lake, Kolkata - 700098, India. \\ Email: chakraba@bose.res.in}

\IEEEauthorblockA{\IEEEauthorrefmark{4} Vikram Sarabhai Space Centre, VRC, ISRO PO, Thiruvananthapuram -
695022, India.}

\IEEEauthorblockA{\IEEEauthorrefmark{5} Indian Space Research Organization - HQ, New BEL Road, Bangalore -
560231, India.}

\IEEEauthorblockA{\IEEEauthorrefmark{6} Moscow Engineering Physics Institute, Moscow, Russia.}} 

\shorttitle{Nandi \etal RT-2 Experiment}

\maketitle

\begin{abstract}

RT-2 Experiment (RT - Roentgen Telescope) is a low energy gamma-ray instrument which is designed and 
developed as a part of Indo-Russian collaborative project of CORONAS-PHOTON Mission to study the Solar 
flares in wide energy band of electromagnetic spectrum ranging from UV to high-energy $\gamma$-rays 
($\sim$2000 MeV). 

RT-2 instruments will cover the energy range of 15 keV to 150 keV extendable up to $\sim$1 MeV. It 
consists of three detectors (two Phoswich detectors, namely, RT-2/S, RT-2/G and one 
solid-state imaging detector RT-2/CZT) and one processing electronic device (RT-2/E). 
Both Phoswich detectors will have
time resolved spectrum, whereas the solid-state imaging detector will have high resolved image of the
solar flares in hard X-rays. We have used Co-57 (122 keV) radio-active source for onboard calibration of
all three detectors. In this paper, we briefly discuss the in-flight performance of RT-2 instruments and
present initial flight data from the instruments. 

This mission was launched into polar LEO (Low Earth Orbit) ($\sim$550 km) on 30th January 2009 from 
Plesetsk Cosmodrome, Russia.

\end{abstract}
%\IEEEpeerreviewmaketitle

%\begin{IEEEkeywords}
% CORONAS-PHOTON, solar flares, gamma-ray spectrometer
%\end{IEEEkeywords}

\section{Introduction}
 
RT-2 Experiment (RT - Roentgen Telescope) is one of the primary instruments of KORONAS-FOTON (Russian) mission
\cite{IEEEhowto:corpho}, also known as CORONAS-PHOTON, in the low energy $\gamma$-ray domain to study the energy 
output during solar 
flares and its spectral evolution. The other major instruments are: a high energy $\gamma$-ray spectrometer
NATALYA-2M, a hard X-ray polarimeter PENGUIN, a ultra-violet imager TESIS etc. All these instruments 
are pointed to the SUN to gather scientific data during solar flares in the energy
band of UV to high energy $\gamma$-ray radiation.
 
The RT-2 Experiment \cite{IEEEhowto:techmod} is comprised of three detector payloads, namely RT-2/S, RT-2/G 
(both Scintillator/Phoswich detectors), RT-2/CZT (solid state detector) and one processing electronics 
device RT-2/E. 

The Phoswich detector payload houses the low energy gamma ray / hard X-ray detector system and front-end 
electronics. The RT-2/S and RT-2/G detector assembly consist of NaI(Tl) / CsI(Na) scintillator in 
phoswich assembly viewed by a photomultiplier tube (PMT) which is the central part of the RT-2/S 
(RT-2/G) detector system. This entire assembly as per design requirement is procured from M/S Scionix 
Holland BV, The Netherlands. Both the detector assemblies sit behind a mechanical slat collimator 
surrounded by a uniform shield of Tantalum material and having different viewing angles of 4$^\circ$ x 
4$^\circ$ (RT-2/S) and 6$^\circ$ x 6$^\circ$ (RT-2/G). As per design and scientific requirement, RT-2/S 
will work in the energy range of 15 keV to 150 keV, extendable up to 1 MeV, whereas aluminum filter is 
used to cut-off low energy photons ($\leq$ 20 keV) for RT-2/G payload. The effective area of each detector
is $\sim$100 cm$^2$ with an average energy resolution of $\sim$18$\%$ @60 keV.

The RT-2/CZT consists of three CZT detector modules (OMS40G256, procured from Orbotech Medical Solutions 
Ltd., Israel) and one CMOS detector (RadEye-1, Rad-icon Imaging Corp., USA) arranged in a 2 x 2 array. Each 
module of CZT detector consists of 256 individual pixels (detectors) of 2.5 mm x 2.5 mm, 
which are controlled by 2 ASIC 
and one CMOS detector consists of 512 x 512 pixels of individual pixel dimension of 50 $\mu$m. The entire 
CZT-CMOS detector assembly sits behind a collimator with two different types of coding devices, namely 
Coded Aperture Mask (CAM) and Fresnel Zone Plate (FZP), surrounded by a uniform shield of Tantalum material 
and has varying viewing angle of 6$'$ - 8$^\circ$. RT-2/CZT payload is the only imaging device in the 
CORONAS-PHOTON mission to image the solar flares in hard X-rays of energy range 20 keV to 150 keV. The
effective area of 3 CZT modules is 48 cm$^2$ with an average energy resolution of 8$\%$ @60 keV 
(at 10$^\circ$C). All three CZT detectors have the spectral information along with high resolution imaging
capabilities. On the other hand, CMOS detector has an effective area of 4.5 cm$^2$ with high resolution
imaging capability only. 

All three payloads are calibrated in flight using Co-57 (122 keV) source of strength 100 nC 
(maximum). A pellet of this source is embedded into one of the slats of the collimators. The detector 
characteristics and specifications are given in the following table-1.

The RT-2 detector system weighs about 55 kg (maxm.) with 15 kg of each detector payload and 10 kg of 
electronic processing device. All three-detector systems (RT-2/S, RT-2/G \& RT-2/CZT) are interfaced with 
the Satellite system called SSRNI through RT-2 Electronic processing payload (RT-2/E). RT-2/E receives necessary 
commands from the satellite system and passes it to the individual detector system for proper functioning
of the detector units. 
%and acquires data from the detector system and stores in its memory for further processing. 
A maximum of 10 MB memory size is allocated for RT-2 data. Data is downloaded twice in a day 
from the Satellite system depending on the radio communication with the ground station.

\begin{table}[!t]
\renewcommand{\arraystretch}{1.3}
\caption{Detector characteristics \& specifications}
\label{First Table}
\centering
\begin{tabular}{|c|c|c|}
\hline
{\bf Payload} & RT-2/S \& RT-2/G & RT-2/CZT \\
\hline
{\bf Detector Type} & NaI(Tl) + CsI(Na) & CZT, CMOS \\
\hline
{\bf Thickness (mm)} & 3 + 25 & 5, 3 \\
\hline
{\bf Size (mm)} & 116 dia & 40 x 40, 24.5 x 24.5 \\
\hline
{\bf Readout} & PMT & Pixels \\
\hline
{\bf Effective area (cm$^2$)} & 100 & 48, 4.5 \\
\hline
{\bf Energy resolution} & 18\% & 8\%, Nil \\
\hline
{\bf Energy range} & 15 - 150 keV (S) & 20 - 150 keV (czt)  \\
                 & 25 - 150 keV (G) & 20 - 150 keV (cmos)     \\              
\hline
{\bf Time resolution} & 1 sec + 100 sec spec & 1 sec \\
		    & 100 msec + 10 sec spec & 100 sec \\
		    & 10 msec + 1 sec spec &  \\
\hline
\end{tabular}
\end{table}

\section{Overall Performance}

\subsection{Orbit-Temperature Profile}

RT-2 instruments onboard CORONAS-PHOTON mission were switched ON during Orbit No. 304 on 19th February 
2009 (16:41:01.919 UT). Before switching ON, the orbit temperatures of the payloads were maintained at around 
16$^\circ$C for RT-2/S \& RT-2/G and -15$^\circ$C for RT-2/CZT. 
\begin{figure}[h]
\centering
\includegraphics[width=1.95in,angle=-90]{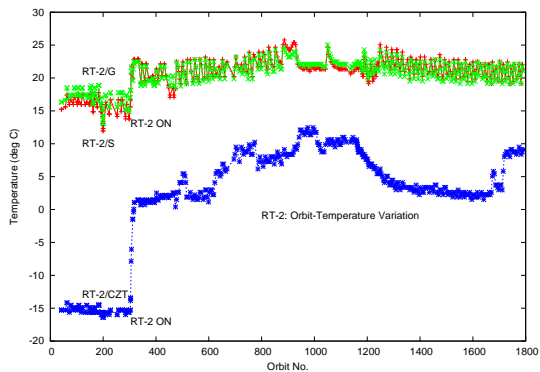}
\caption{Temperature variation of the RT-2 instruments in different orbits.}
\label{orb-temp}
\end{figure}
After switching ON, the orbit temperature 
variation (payload temperature) was found to be in comfortable range of 18$^\circ$C to 25$^\circ$C 
for RT-2/S \& RT-2/G and 2$^\circ$C to 11$^\circ$C for RT-2/CZT. The orbit-temperature profile is 
shown in Figure 1. The fluctuation in the temperature profile is due to the High Voltage (HV) operation 
of each payload and temperature variation in each orbital condition (space environment).

\subsection{Charge Particle Variation}

The CORONAS-PHOTON satellite orbits the Earth at a altitude of $\sim$550 km and an inclination of 
$\sim$82.5$^\circ$. The LEO and high inclination affects the satellite GOOD time observation as it passes through
the South Atlantic Anomaly (SAA), North Pole (NP) and South Pole (SP) regions. Due to this constraints, we
could have around 40\% good time interval data in each orbit.
During initial phase of operation, RT-2 instruments were kept throughout at lower high voltage (HV) till 
all issues related GOOD/BAD command operation were settled. 

\begin{figure}[h]
\centering
\includegraphics[width=1.95in,angle=-90]{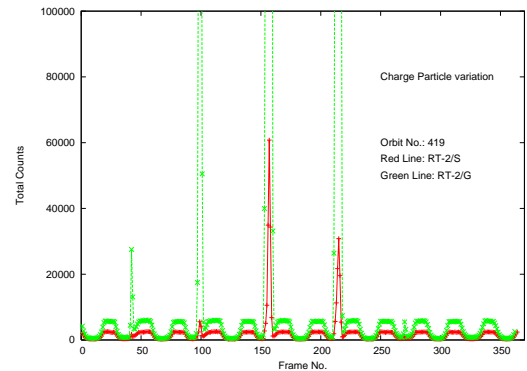}
\caption{Photon Count variation of RT-2 detectors due to charge particles of SP, NP and SAA region.}
\label{count-cp}
\end{figure}

In this condition, the variation of photon counts as registered by the RT-2 detectors, while passing 
through the high background region (SP, NP and SAA), are shown in Figure 2. It is clear that
that the periodic variation in photon counts (per 100 sec) is repeating in exactly 95 to 100 minutes of
period which is one complete orbital time period of the satellite.   
The maximum count rate (25 cts/sec for RT-2/S and 57 cts/sec for RT-2/G) variations show a normal 
behavior of characteristics of SP and NP. Both the detectors also registered huge counts while passing 
through the SAA region. RT-2/CZT detector modules also registered the particle events with maximum 
rate of 2 cts/sec. During this observation (Orbit No. 419), all health parameters of the instruments 
were normal.

\subsection{Instrument Operation \& Health Status}

The operational aspects of both RT-2/S and RT-2/G payloads are identical and they operate using a supply of
27$^{+7}$$_{-3}$ Volt. 
The total power consumption is limited to 4.5$\pm$0.5 Watt for each payload. The input power is converted 
to +15 Volt and +5 Volt with the help of a low voltage MDI DC-DC converter for the necessary requirement 
to drive different components of the individual payloads. The +15 Volt is also converted separately through 
a voltage regulator circuit to high voltage ($\sim$700 Volt) needed for biasing the PMT. The PMT is operable 
in the range of 400 Volt to 900 Volt and change in HV is commandable in $\sim$4.5 Volt increments. The Pulse 
Shape Discriminator (PSD) from two crystals (NaI \& CsI) and Lower Level Discriminator (LLD) for two 
amplifiers (G1 \& G2) are also commandable. Voltage Control Oscillator (VCO) is used to monitor the 
instrument health parameters, such as +5 Volt supply, Temperature variation (Thermistor), High Voltage (HV) 
and LLD control.  Till now, the RT-2/S and RT-2/G detectors are working normally.

In-flight condition, high voltage (HV) of both Phoswich detectors was increased in steps to check the linearity 
of HV values with the VCO counts. 
%Different HV commands were sent to the detectors and in each HV operation, 
%the detector functionality and health parameters were verified. 
At present, RT-2/S \& RT-2/G are operated 
with HV command of 07B2 (763 Volt) and 27AA (727 Volt) respectively and both the Phoswich detectors were 
working normally. The linearity plot for HV calibration of RT-2/S is shown in Figure 3. It is also noted 
that the lower cut-off of HV for RT-2/S is around 310 Volt.

\begin{figure}[h]
\centering
\includegraphics[width=1.95in,angle=270]{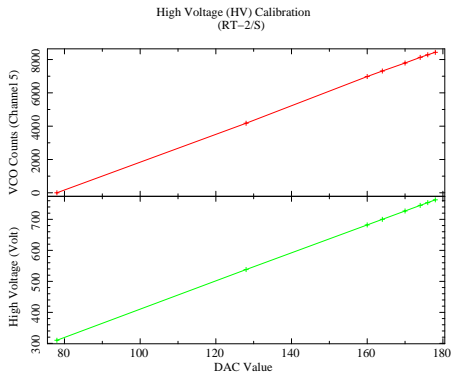}
\caption{Linearity calibration of HV operation of RT-2/S detector.}
\label{hv-cal}
\end{figure}

RT-2/CZT payload \cite{IEEEhowto:malkar} also operates with 27$^{+7}$$_{-3}$ Volt. The total power 
consumption is limited to 7.5$\pm$ 0.5 Watt. 
The input power is converted to +15 V and +5 V with the help of a MDI DC-DC unit for the required supply 
of different components of the instrument. A non-controllable high voltage generator is used to bias (fixed) 
the CZT detector with -600 Volt.

An Analog-Digital Converter (ADC) instead of VCO is used to monitor the instrument health of RT-2/CZT 
payload. The ADC output monitors +5 Volt supply, Temperature variation (Thermistor), High Voltage (HV) 
OFF/ON, CZT detector supply and CMOS supply. Till now, the RT-2/CZT detector is working normally.

\section{Phoswich Detectors}

Presently, RT-2/S and RT-2/G are operated with proper HV. The channel spectra of NaI and Pulse Shape 
Discriminator (PSD) spectra of both the detectors are normal and Pulse Height (PH) is stable throughout 
the present orbital operation. The channel spectra of both the detectors are controlled by two 
post-amplifiers (G1 and G2). The G1 spectrum is sub-divided in two parts of NaI-G1 and CsI-G1 spectra
depending on the pulse output from the crystals. The NaI-G1 and CsI-G1 spectra are calibrated with the
working energy ranges from 15 keV to $\sim$100 keV and 30 keV to $\sim$170 keV, whereas G2 will work in the 
energy range of 100 keV to $\sim$1 MeV. 
%Instead of long duration background data, we have presented only
%a small segment of 100 sec of observation.  

\subsection {Spectrum (RT-2/S)}

The PSD and channel spectra of NaI crystal of RT-2/S \cite{IEEEhowto:debnath} are shown in Figure 4. 
RT-2/S is operated with HV of 763 Volt (CMD 07B2h). PSD spectrum shows the separation between the NaI and 
CsI events. The Pulse Shape (PS) is applied at 26 channels (051Ah) to discriminate the NaI and CsI events. 
The PH around 540 Channel of NaI spectrum is the signature of background peak due to the decay of activated 
I$^{121}$ atom. 
%The 122 keV (Co-57) line is observed in G2 spectrum (not shown).

%\begin{figure}[h]
%\centering
%\includegraphics[width=2.0in,angle=0]{Orb1144-27AA-Fr249-G-NaI.ps}
%\caption{NaI Spectrum of RT-2/G}
%\label{nai-spec-g}
%\end{figure}

%\begin{figure*}[!t]
\begin{figure}[h]
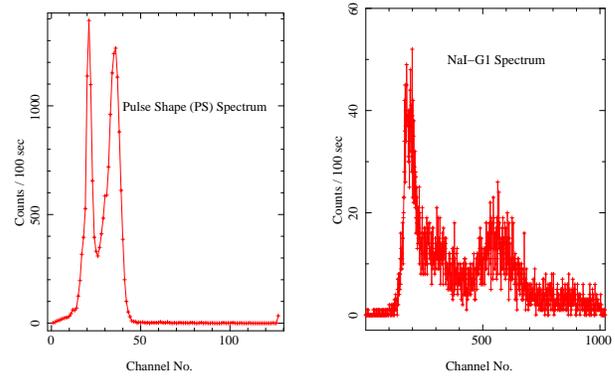

%\centerline{\subfloat[Case I]\includegraphics[width=2.5in]{Orb1144-27AA-Fr249-G-NaI.ps}
\centerline{
\includegraphics[width=1.5in,angle=0]{orb1661-S-psd-107fr.ps}
\label{fig_first_case}
\hfil
\includegraphics[width=1.5in,angle=0]{orb1661-S-NaI-107fr.ps}
\label{fig_second_case}}
\caption{RT-2/S Spectrum: PSD and NaI(Tl) spectra (100 sec observed data).}
\label{s-psd-nai}
\end{figure}

\subsection {Light curve of GRB 090618}

For the first time, since its operation, RT-2 payloads have detected hard X-ray signature of GRB 090618.
Light curve of GRB 090618 with 1 sec time resolution is shown in Figure 5. The multi-peak profile of the 
light curve of GRB 090618 also detected by other satellites (SWIFT, KONUS-RF, Fermi etc.) 

\begin{figure}[h]
\centering
\includegraphics[width=1.95in,angle=270]{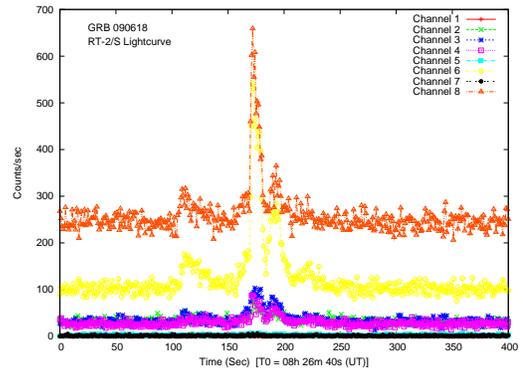}
\caption{Light curve of GRB 090618 as detected by RT-2/S.}
\label{s-lc-grb}
\end{figure}

\subsection {Spectrum (RT-2/G)}

RT-2/G detector is operated with HV of 727 Volt (27AAh). The Channel spectra of PSD and
NaI are shown in Figure 6. The Pulse Shape (PS) is applied at 28 Channel (251Ch) to 
discriminate the NaI and CsI events in the detector. The PH around 590 Channel of NaI spectrum is the 
signature of background peak due to the decay of activated I$^{121}$ atom. Both the PS and PH are stable 
throughout the 1661 orbit operation. 
%The 122 keV (Co-57) line is observed in G2 spectrum (not shown).  
From the Channel-energy calibration, it is identified that the background peak (I$^{121}$) is detected 
at $\sim$58.0 keV with energy resolution $\sim$30\%. 
%The decrease in energy resolution is due to the weakness of the source (Co-57).

%\begin{figure}[h]
%\centering
%\includegraphics[width=2.0in,angle=0]{Orb1144-27AA-Fr249-G-NaI.ps}
%\caption{NaI Spectrum of RT-2/G}
%\label{nai-spec-g}
%\end{figure}

%\begin{figure*}[!t]
\begin{figure}[h]
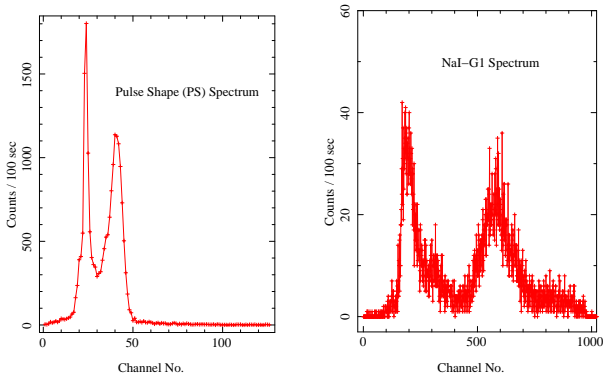

%\centerline{\subfloat[Case I]\includegraphics[width=2.5in]{Orb1144-27AA-Fr249-G-NaI.ps}
\centerline{
\includegraphics[width=1.5in,angle=0]{orb1661-G-psd-107fr.ps}
\label{fig_first_case}
\hfil
\includegraphics[width=1.5in,angle=0]{orb1661-G-NaI-107fr.ps}
\label{fig_second_case}}
\caption{RT-2/G Spectrum: PSD and NaI(Tl) spectra (100 sec observed data).}
\label{s-psd-nai}
\end{figure}

\section{Solid-State Detector}

RT-2/CZT payload consists of 3 CZT modules and 1 CMOS detector. CZT modules are operated with -600 Volt and 
CMOS is operated with normal 5 Volt supply. During 441 orbit, CZT detectors were operated with HV -600 volt 
for the first time with threshold 30 keV. After
analyzing the data of all 3 CZT modules, it is observed that calibration source peak (Co-57) is stable 
throughout the orbit. In this paper, we are presenting only CZT2 module data. The spectrum and image \cite
{IEEEhowto:kotoch} are shown in the following figures. 

As the operation of CMOS is quite different from that of CZT, a careful onboard calibration is going on to find
the background threshold. This threshold should be evaluated during SHADOW mode (away from SUN) and the
threshold value should be set for CMOS before its actual operation.

\subsection {Spectrum of CZT2 module}

The background spectrum of CZT2 module is shown in Figure 7. The calibration source peak of Co-57 (122 keV)
is clearly detected by the CZT2 module. The energy calibrated resolution of the peak is around 5.6\% 
@122 keV. Other two CZT modules also detected the the calibration source peak at around 127 keV and 
124 keV with energy resolution 7.32\% and 5.97\% respectively. It is also noted that the energy spectrum
of all 3 CZT modules are background noise dominated below 40 keV.

\begin{figure}[h]
\centering
\includegraphics[width=1.95in,angle=270]{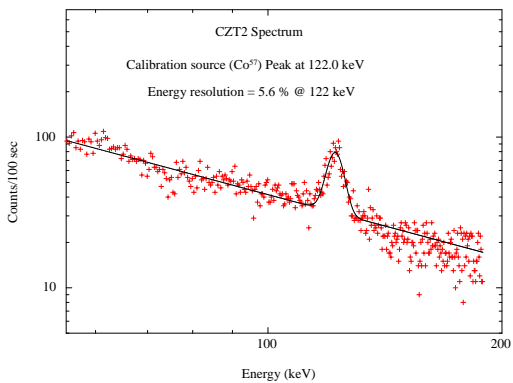}
\caption{Spectrum of CZT2 module (100 sec observed data).}
\label{nai-spec-g}
\end{figure}

The calibration source peak shifted to the higher energy in CZT1 (127 keV) and CZT3 (124 keV) could be due 
to the resultant effect of the noisy pixels. 

\subsection {Image of CZT2 module}

Detailed analysis of all individual pixels (3 x 256 pixels) of each 3 modules reveals that some of
the pixels of CZT1 and CZT3 modules are noisy. 
%So, in later phase of operation the CZT modules, the threshold 
%was changed to 40 keV to understand the noisy behavior of the pixels.  
Even after changing the threshold (40 keV), the noise pattern of few individual pixels remained. 
This could be due to leakage of UV rays through the collimator to the CZT modules or may be due to some unknown
factor. It will take some more time to understand the system completely. 
The background image of CZT2 module is shown in Figure 8 with gray scale distribution of all 256 pixels with 
less count (black color) and high count (white color) values. 

\begin{figure}[h]
\centering
\includegraphics[width=1.95in,angle=0]{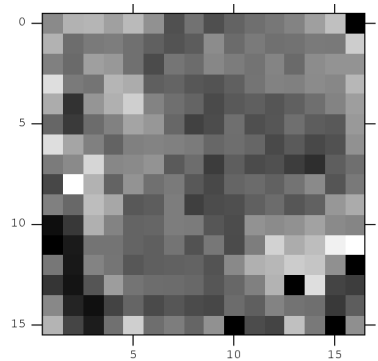}
\caption{Image of CZT2}
\label{img-czt2}
\end{figure}
\section{Conclusion}
All the 3 scientific payloads are working properly with good health parameters. 
In-flight characterization is carried out with the background data (SHADOW mode), as the SUN is not active. 
%The energy ranges of each individual payload are in the expected energy-band to study the solar flares. 
RT-2 payloads have detected GRB 090618 and initial result (light curve) is presented here.
To have a good scientific outcome from RT-2 instruments, one may have to wait for the high energy flare 
($\ge$ 15 keV) to occur in the SUN.

\section*{Acknowledgment}

The authors would like to thank the engineers and technical staffs from DAA/TIFR and from BSED, QDTE of 
VSSC for their kind help towards the successful completion of the project. We would also like to
acknowledge the Indian Space Research Organization (ISRO) for financial support and overall management.

\end{document}